\documentclass[12pt]{article}

\usepackage{epsfig,amsmath,amsfonts,cite}

\newcommand{\reff}[1]{(\ref{#1})}

\newcommand{\sumprime}[2]{\sum_{#1\in{\mathbb{Z}}^{#2}}\!\!'}
\newcommand{\sumnoprime}[2]{\sum_{#1\in{\mathbb{Z}}^{#2}}\!\!}
\newcommand{\mt}[1]{\mathcal{M}[#1]}
\newcommand{\mta}[2]{\mathcal{M}_{#1}[#2]}

\newcommand{\lit}[1]{\mathcal{L}^{\scriptscriptstyle -1}[#1]}

\newcommand{\R}{\text{Re}\;}

\newcommand{\Imint}[1]%
 {\frac{1}{2\pi\, i}\,\int_{c-i\infty,\; c#1}^{c+i\infty}\!\!\!ds\ }

\newcommand{\fsub}{f_{\text{sub}}}

\begin{document}


\begin{flushright} CERN--TH/2001--363 \end{flushright}
\vspace{5mm}
\vspace{0.5cm}
\begin{center}

{\Large \bf On dimensional regularization of sums} \\[1cm]
{\large Roberto Contino$^{1,2}$, Andrea Gambassi$^1$}
\\[1.5cm]

{\small
$^1$\textit{Scuola Normale Superiore $\&$ INFN, Piazza dei Cavalieri 7, 
 I-56126 Pisa, Italy}  \\[0.3cm]
$^2$\textit{Theory Division, CERN, CH-1211 Geneva 23, Switzerland} }

\end{center}

\vspace{1cm}


\hrule \vspace{0.3cm}
{\small  \noindent \textbf{Abstract} \\[0.3cm]
\noindent
We discuss a systematic way to dimensionally regularize divergent sums arising
in field theories with an arbitrary number of physical compact dimensions or 
finite temperature.
The method preserves the same symmetries of the action as the conventional
dimensional regularization and allows an easy separation of the regulated 
divergence from the finite term that 
depends on the compactification radius (temperature).

\vspace{0.5cm}  \hrule



\section{The problem}

In a variety of problems one has to deal with formally divergent sums, usually related to
Feynman diagrams with one or more discrete momenta, as in the case of theories with compact
extra-dimension, finite-size scaling theory in critical phenomena, thermal field theory.
It is crucial to find a regulator that 
preserves the symmetries of the problem and leads to a simple computational procedure.
When the momenta are not discrete but continuous, such a procedure exists and it is the 
well-known dimensional regularization (DR) of integrals~\cite{DR,Wilson}.
In this paper we discuss a systematic way to obtain the dimensional regularization of 
an important class of sums, following closely the analogy with the case of a continuous 
variable~\cite{Collins}.
Even if practical recipes to deal with particular examples have been given in the past
and the use of special functions to this purpose is not 
new~\cite{statisticalM,finiteT,KK, effCas,zfunction,Ogreid},
our aim is to provide a general method, which extends and in a sense justifies
the analysis of ref.~\cite{GravitonLoops}, where the idea of dimensionally regularized
series was applied to a number of different loop sums.
In~\cite{Nibbelink} the approach to dimensional continuation was adopted and combined with 
complex analysis techniques to delineate a general procedure,
restricting however to the case of only one physical compact dimension. 
In~\cite{Bedingham} the use of the Mellin transform, together with 
dimensional regularization was introduced to deal with asymptotic expansion 
of series in thermal field theory, formally the same problem as discussed in~\cite{Nibbelink}.  

We propose to {\it define} sums in complex dimension using the analytic properties
of a generalized zeta function,
resulting in a simple method where the regulated divergence can be easily separated from the
finite part.
Our technique has some common points with the well known zeta-function
regularization~\cite{zfunction},
which leads to quite similar calculations but is well distant from the spirit of analytical
continuation in the number of dimensions.

We hope that this work may contribute to clarify some debated aspects of regularization
in extra-dimensional models, where it is crucial to preserve the symmetries of the action
\cite{BHN,NG,Quiros,CP}.
In \cite{NG} it was suggested that the finite result obtained 
by Barbieri, Hall and Nomura~\cite{BHN}  for the radiative correction to the Higgs mass 
coming from the Yukawa sector of their model was a regularization artifact.
Unfortunately, the authors of ref.~\cite{NG} made use of a sharp cut-off on the series,
which explicitly breaks the supersymmetry of the model of ref.~\cite{BHN} and invalidates
their argument. 
The result of \cite{BHN} was reobtained in~\cite{Quiros} by using a thick brane as a 
regulator for the series and in~\cite{CP} in two different ways, with a Pauli--Villars and 
by using dimensional regularization.   
In this work we give a formal procedure to introduce the dimensional
regularization of a series and demonstrate in detail some important properties.
In particular, we show that there is {\it no ambiguity} in exchanging the series 
and the integral over loop momenta if both are properly regularized.

The paper is organized as follows.
In Section~\ref{definition} we briefly review standard dimensional
regularization of integrals and give our rules for extending it to the case of a series
by using a generalized zeta function.
We discuss the infinite radius limit in Section~\ref{finitepart},
showing that the (regulated) divergence equals
that of the corresponding integral and as such does not depend on the radius.
A representation using theta-functions is also given in Section~\ref{threp},
which is useful to perform
explicit computations of sums in a class of physical problems.
In Section~\ref{example} we show how the method works on an explicit example,
computing the Casimir energy of a massive scalar field. 
In Section~\ref{conclusions} we draw our conclusions.

Finally, in the appendices we briefly recall the definition of the Mellin 
transform~(\ref{appmellin}), present the analytic continuation of the generalized zeta 
function, together with its asymptotic behaviour~(\ref{Appzeta}) and  
discuss some important properties of the dimensionally regularized 
series~(\ref{properties}).

\section{Dimensionally regularized series}
\label{definition}

Dimensional regularization of integrals was introduced in~\cite{DR} as a simple 
tool to manage the divergences that arise in (perturbative)
field theory, preserving the gauge symmetry. It was later
derived as an axiomatic procedure by Wilson~\cite{Wilson}, see~\cite{Collins} for a
detailed discussion.
The problem is to give a meaning to the integration of a function
$f(p^2)$ over a space of complex dimension $d$, getting the usual result whenever $d$ 
is an integer and the integral exists in the ordinary sense.
Without going into too much detail, one can {\it define} the integral through the formula
\begin{equation} \label{eq:integraldefinition}
\int\!\!d^dp\, f(p^2) = \Omega_d \int_0^{\infty}\!\!dp\, p^{d-1}\, f(p^2) \; ; \qquad
 \Omega_d = \frac{2\pi^{d/2}}{\Gamma(d/2)}\quad ,
\end{equation}
for all (complex) values of $d$ for which the integral converges and then analytically 
continue the result to the desired value.
Typically one has to cope with ultraviolet (UV) divergences, which are cured by 
considering a sufficiently small (real part of) $d$ and appear as 
poles in the final expression for the integral.
It may happen that the value of $d$ that makes~\reff{eq:integraldefinition}
UV-convergent is so small that the integral diverges in the infrared (IR):
in this case one defines the integral subtracting the leading behaviour of the function
$f$ for $p^2\to 0$. 
Let us suppose that $f(t)$ has the following asymptotic expansion for $t\to 0$:
\begin{equation}
\begin{gathered}
f(t) = a_0\,t^{\alpha_0} + a_1\,t^{\alpha_1} +
 a_2\,t^{\alpha_2} + \cdots \\
\alpha_0 < \alpha_1 < \alpha_2 < \cdots
\end{gathered}
\label{eq:andam}
\end{equation}
and $f(t)\sim t^{-\rho}$ for $t\to\infty$.
Then, for $-\alpha_q<\R d/2<\min (-\alpha_{q-1},\rho)$, $q$ a positive integer, the integral
is defined by
\begin{equation} \label{eq:subtracteddefinition}
\int\!\!d^dp\, f(p^2) = \Omega_d \int_0^{\infty}\!\!dp\, p^{d-1}\,
 \left[ f(p^2) - \sum_{k=0}^{q-1} a_k (p^2)^{\alpha_k} \right] \quad .
\end{equation}
If $-\alpha_0<\rho$ subtractions are not really needed and the integral on the r.h.s. 
of eq.~\reff{eq:integraldefinition} converges for $-\alpha_0<\R d/2<\rho$; in this case 
eq.~\reff{eq:subtracteddefinition} is simply the correct analytic
extension to the interval $-\alpha_q<\R d/2<-\alpha_{q-1}$, $q>0$. On the contrary, when
$-\alpha_0>\rho$, there is no value of $d$ that makes~\reff{eq:integraldefinition}
both UV- and IR-convergent and~\reff{eq:subtracteddefinition} becomes a definition.

This general procedure does not work in the case of sums, given that there is no closed form 
for what corresponds to the ``solid angle'' $\Omega_d$\ in the case of a hypercubic 
lattice and one cannot directly take advantage of the spherical symmetry of the function 
$f$. We then proceed as described below, following closely the analysis of~\cite{Collins}
for the integrals.

We want to give a meaning to the expression
\begin{equation}
\sumprime{n}{d} f(n)\quad ,
\label{somma}
\end{equation}
where the prime means that $n=0$ is omitted in the sum and $f$\ is assumed to be continuous.
We restrict for simplicity to the case of a scalar function (the
case of tensorial functions can be addressed in the same way as for standard
DR of integrals) and require covariance of our result under (discrete) rotations of the
hypercubic lattice. In the case of only one variable, this implies that the function
$f$ depends only on the norm of the vector $n$ and not on its direction.
The general case in which $f$ depends also on external momenta is considered below.

An operation of summation in arbitrary complex dimension is uniquely determined
\footnote{Except for an arbitrary normalization, which can be fixed on a set of basis 
functions~\cite{Wilson,Collins}.
We require the usual result of integer dimensions 
 \begin{equation*}
  \sum_{n\in \mathbb{Z}^d} e^{-\pi s n^2} = \vartheta^d (s) \quad ,
 \end{equation*}
to hold for all $d\in\mathbb{C}$ (see the appendix for the definition of the 
$\vartheta$-function).
 }
by requiring the following basic properties, valid for standard summation:
\begin{enumerate}
\item {\it Linearity}: for any complex numbers $a,b$
 \begin{equation} \label{eq:axiom1}
  \sumprime{n}{d} \left[ a\ f(n) + b\ g(n) \right] =  a \sumprime{n}{d} f(n) +
  b \sumprime{n}{d} g(n) \quad ;
 \end{equation}
\item {\it Invariance under lattice translations}: for any vector $q$
 \begin{equation} \label{eq:axiom2}
 \sumnoprime{n}{d} f(n+q) =  \sumnoprime{n}{d} f(n)\quad .
\end{equation}
\end{enumerate}

Notice that the scaling axiom~\footnote{It states that $\forall\, s >0$,
 \begin{equation*}
  \int\!\!d^dp\, f(s^2 p^2) = s^{-d}\int\!\!d^dp\, f(p^2)\quad .
 \end{equation*}
See~\cite{Collins}.
} required for dimensionally regularized integrals does not hold for the series.
As for the standard case of integrals, vectors are thought to lie in an infinite
dimensional space, with the difference that now each component of the vector 
has an integer value.
The dimensionality $d$ is introduced by the sum operation with the requirement that if
$d$ is a positive integer all vectors collapse in a $d$-dimensional subspace.
When the function $f$ depends also on some external momenta $q_i$
(actually only through the scalar products $(n\cdot q_i)$, $(q_i\cdot q_j)$
being a scalar function), one can proceed again in complete analogy with~\cite{Collins}:
it is always possible to find an $N$-dimensional sublattice $\mathbb{Z}^N$
(with $N$ finite being the external vectors in finite number),
which contains all the external vectors.
Let us decompose $n$ into a longitudinal and a transverse part with respect 
to $\mathbb{Z}^N$: $n=n_\parallel + n_{\perp}$, so that
\begin{equation} \label{eq:defgen}
\sumnoprime{n}{d} f(n,q_i) = \sum_{n_\parallel\in\mathbb{Z}^N} 
 \sumnoprime{n_\perp}{d-N}
 f(n_\perp^2+n_\parallel^2,n_\parallel q_i,q_i\cdot q_j) \quad .
\end{equation}
The outer sum in the r.h.s. of~\reff{eq:defgen} is a standard series on the lattice
$\mathbb{Z}^N$, while the inner one can be defined through~\reff{somma}, $f$ now being
independent of the direction of $n_\perp$.
Then there is no loss of generality in reducing to the case in which $f$ is a function
of only the dummy variable.

Let us now describe an explicit procedure for summing in complex dimensions.
Given a function $f(t)$ continuous for $t\in\mathbb{R}^+$, let us assume that
$\theta(t-a)f(t)$ (where $\theta$\ is the step function) is Mellin-transformable with fixed
$0< a < 1$ (see Appendix~\ref{appmellin} and Ref.~\cite{Davies,GraRyz} for details 
on the Mellin transform), i.e. $\exists\; \rho \in {\mathbb R}$ such that
\begin{equation*}
\int_a^{\infty} \!\! dt\ |f(t)|\ t^{\sigma-1}\  <\  
 \infty \qquad \forall\, \sigma < \rho \quad .
\end{equation*}
Then the Mellin transform
\begin{equation}
\mta{a}{f,s} \equiv \mt{\theta(t-a)f(t),s} = \int_a^\infty\!\! dt\ f(t)\ t^{s-1}
\label{MelTrans}
\end{equation}
is an analytic function for $s\in\mathbb{C}$\ if $\R s< \rho $.
In particular, the assumption of the existence of the Mellin transform excludes
from our discussion those functions $f(t)$\ growing exponentially for 
$t\rightarrow +\infty$.
The inversion theorem guarantees that
\begin{equation}
\theta(t-a)f(t)\ =\ \Imint{<\rho} \mta{a}{f,s}\ t^{-s} \quad .
\label{eq:antiM}
\end{equation}
Usually, the Mellin transform of the function $f$ is defined in the strip
$-\alpha_0<\R s<\rho$, if $f(t)\sim t^{\alpha_0}$
for $t\to 0$; choosing the parameter $a>0$ allows the lower limit $-\alpha_0$ to be sent 
to $-\infty$.
Let us now assume that there exists an integer value of $d$\ such that~\reff{somma}
exists in the ordinary sense. It is easy to realize that if it is the case, then 
$d/2 < \rho$ and
\begin{equation}
\sumprime{n}{d} f(n^2) = \Imint{<\rho} \mta{a}{f,s} \sumprime{n}{d}\ \frac{1}{(n^2)^s}\quad ,
\end{equation}
where exchanging the series with the integral is allowed in the domain of uniform 
convergence (see below).
We define a generalized $\zeta$-function (a particular Epstein's zeta function, 
see~\cite{zfunction} and~\cite{luscher})
\begin{equation}
  \zeta(s,d) \equiv \sumprime{n}{d}\ \frac{1}{(n^2)^s}    \quad ,
\end{equation}
whose properties are discussed in Appendix~\ref{Appzeta}.
This series converges uniformly in any closed subset of the line $\R s > d/2$,
i.e. the function $\zeta(s,d)$\ is analytic in the half-plane $\R s > d/2$.
Thus we may give the following representation of our original sum:
\begin{equation}
\sumprime{n}{d} f(n^2) = \frac{1}{2\pi\, i}\,
 \int_\Gamma \!\!ds\ \zeta(s,d) \mta{a}{f,s} \; ; \qquad
 \Gamma=\{ \R s=c; d/2<c<\rho \}\quad .
\label{repres}
\end{equation}
Let us observe that both $\mta{a}{f,s}$\ and $\zeta(s,d)$\ can be analytically continued
outside their definition domains. Their analytic continuations will generally have singularities
in the complex plane. Now the recipe to define the sum~\reff{somma}\ even for $d/2\ge\rho$ is
clear: it is simply the continuation of integral~\reff{repres}, where we consider the analytic
continuation of $\mta{a}{f,s}$\ and $\zeta(s,d)$\ in the integrand.
In the same way we can {\it define}~\reff{somma}\ by using the representation~\reff{repres}\
even if there is no integer value of $d$\ such that $d/2<\rho$ (for example if
$f(x)\sim x^{\alpha}$\ for $x\rightarrow\infty$\ and $\alpha\ge -1$) and the same considerations
apply for complex $d$ if one considers $\R d$\ instead of $d$ in previous relations.

When $\R d$ is increased towards values greater than $\rho$, the integral~\reff{repres} gets a
residue contribution from the pole of $\zeta(s,d)$ in $s=d/2$: as we will show below,
this term is divergent 
for $\R d/2>\rho$ and coincides with the infinite radius limit
\footnote{If the physical value of $d$ is less than $2\rho$ the series is convergent and 
there is no need to (dimensionally) regularize it; in this case the representation~\reff{repres}
is still useful for an explicit evaluation. }.
The remaining complex integral along the contour $\Gamma=\{ \R s=c; c<\rho<d/2 \}$
corresponds to the finite radius-dependent part; in the following section we give a
practical recipe to evaluate it using the $\vartheta$-function.
Note that the operation~\reff{repres} respects the required properties:
linearity follows from linearity of the Mellin transform; translational invariance follows
from translational invariance of the usual sum over the transverse subspace, if this is
taken sufficiently large to contain the vector $q$ in eq.~\reff{eq:axiom2}.
In Appendix~\ref{properties} we prove some remarkable properties which hold for
the sum over complex dimensions.

Finally, a comment on~\reff{repres} is in order to clarify the meaning of the parameter $a$:
as already said after eq.~\reff{eq:integraldefinition}, going to too
small $d$ without performing an appropriate subtraction in the case of the integral
would introduce a spurious IR divergence which has no physical meaning.
In the same way, had we not introduced the parameter $a>0$, the Mellin transform
\begin{equation*}
\int_0^\infty\!\! dt\ f(t)\ t^{s-1}
\end{equation*}
would have been divergent (for any $s$) in all cases in which the integral corresponding 
to the series
needs IR subtraction to be defined in dimensional regularization.
Setting $a>0$ is a simple way of avoiding IR subtractions for the series.
We prove below that the final result does not depend on $a$, as one expects having the
initial series~\reff{somma} no IR problems at all.

\section{An alternative representation}
\label{finitepart}

Once given the definition of a series in complex dimensions, we present an alternative 
representation of~\reff{repres} to show that
\begin{itemize}
 \item[--] the result is independent of the actual value of $0<a<1$;
 \item[--] the UV dimensional poles are the same as those of the corresponding integral;
 \item[--] the property (analogous to that for ordinary sums)
   \begin{equation} \label{eq:infiniteR}
   \lim_{R\to \infty} \frac{1}{R^d} \sumprime{n}{d} f(n^2/R^2) =
   \int d^dp\ f(p^2)
   \end{equation} is formally valid.
\end{itemize}
The final formula will be easier to handle for a numerical evaluation or an expansion
in the parameters.

\subsection{The infinite radius limit}

Our strategy is to obtain first a compact expression for $\R d/2<\rho$
and then analytically continue this result to the case $\R d/2>\rho$. 
Introducing the radius $R$ explicitly, definition~\reff{repres} becomes ($0<a<1$):
\begin{equation}
S(d,R)\equiv\frac{1}{R^d} \sumprime{n}{d} f(n^2/R^2) =  \frac{1}{2\pi i}
 \int_\Gamma ds\ \zeta(s,d)\ R^{2s-d} \mta{a/R^2}{f,s} \quad ,
\end{equation}
where the contour $\Gamma$ has to be fixed according to~\reff{repres}. 
Let us assume that $f(t)$ has the asymptotic expansion~\reff{eq:andam} for $t\to 0$
and  $f(t)\sim t^{-\rho}$ for $t\to\infty$, even if the derivation goes the same way
for more general expansions. We can take 
\begin{equation*}
\Gamma=\{\R s=c\, ;\, -\alpha_q<\R d/2<c<\min(-\alpha_{q-1},\rho) \} \quad ,
\end{equation*}
$q$ being the smallest integer such that $-\alpha_q<\R d/2<\rho$ 
(see Fig.~\ref{fig:contour}).
\begin{figure}
\begin{center}
\epsfig{file=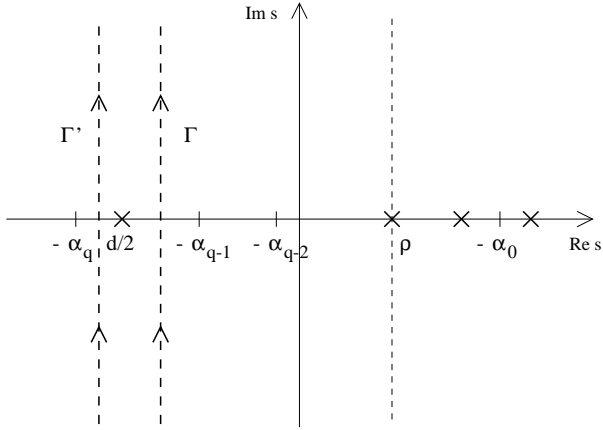,width=8cm}
\caption{{\small Position of the contour $\Gamma$ and $\Gamma'$: 
the poles for $\R s\ge\rho$, are those of the continuation of $\mta{a}{f,s}$.}}
\label{fig:contour}
\end{center}
\end{figure} 
Defining~\footnote{For $q=0$\ the subtraction terms are omitted and we mean 
$\min(-\alpha_{-1},\rho)\equiv\rho$.}
\begin{equation}
f_{\text{sub}}(t) = f(t) - \sum_{k=0}^{q-1} a_k\ t^{\alpha_k} \quad ,
\end{equation}
it follows, for $-\alpha_q<\R s<\min(-\alpha_{q-1},\rho)$:
\begin{equation}
\mta{a/R^2}{f,s} = \mt{\fsub,s} - 
 \int_0^{a/R^2}\!\!dt\, \fsub(t)\, t^{s-1} -\sum_{k=0}^{q-1} 
 \frac{a_k}{s+\alpha_k}\left(\frac{a}{R^2}\right)^{s+\alpha_k} \quad .
\end{equation}
Note that the Mellin transform in the first term is IR-convergent thanks to the 
subtraction made on $f$ and the second term is analytic for $\R s >-\alpha_q$.
Using the asymptotic limit $\zeta(s,d)\to 2d$ for $\R s\to +\infty$
(see Appendix~\ref{Appzeta}),
and moving the contour $\Gamma$ at infinity in the positive half-plane,
one easily finds that the integral in the complex variable $s$ of the second term vanishes, 
since $a<1$. Integrating the third term along $\Gamma$, again closing the contour in 
the positive half-plane, one gets instead a residue contribution
from the poles $s=-\alpha_k$, $k=0,\dots,(q-1)$.
We end up with
\begin{equation} \label{eq:subdef}
S(d,R) = \frac{1}{2\pi i} \int_\Gamma ds\ \zeta(s,d)\ R^{2s-d} \mt{\fsub,s} +
 \sum_{k=0}^{q-1}  \frac{a_k}{R^{2\alpha_k+d}} \zeta(-\alpha_k,d) \quad .
\end{equation}
This expression represents a definition of the series 
for $-\alpha_q<\R d/2<\min(-\alpha_{q-1},\rho)$ alternative to~\reff{repres},
where the use of the IR cut-off $a$ has been replaced by an appropriate subtraction of the
initial function, much in the same way as one does in standard DR.

If $-\alpha_0<\rho$ the IR subtractions are not really required to define the series:
choosing $d$ in the strip $-\alpha_0<\R d/2<\rho$ ($q=0$), eq.~\reff{eq:subdef} 
coincides with~\reff{repres} with $a=0$.
In this case eq.~\reff{eq:subdef} with $q>0$ represents the correct analytical 
extension of the series
to the interval $-\alpha_q<\R d/2<-\alpha_{q-1}$, but it becomes non-trivial
when $-\alpha_0>\rho$ and the subtractions are really necessary to give a meaning 
to the expression.
It is interesting to notice that $\mt{\fsub,s}$ in the first term is nothing but
the dimensionally regularized integral of $f$ in $d=2s$ dimensions, up to a 
solid-angle factor.

The limit $R\to\infty$ cannot be extracted immediately from~\reff{eq:subdef}, 
as one must first extrapolate the result to values of $\R d/2$ greater than $-\alpha_0$.
To analytically continue~\reff{eq:subdef} to $\R d/2>\min(-\alpha_{q-1},\rho)$ we have to 
take into account that increasing $\R d/2$, the pole $s=d/2$ of $\zeta(s,d)$ crosses
the contour $\Gamma$ and gives a residue term. To simplify the analytic continuation,
we can rewrite~\reff{eq:subdef}\ in a suitable form,  evaluating this residue term from 
the very beginning, i.e. moving the contour $\Gamma$\ into a new one 
(see Fig.~\ref{fig:contour})
\begin{equation*}
\Gamma^\prime=\{\R s=c\, ;\, -\alpha_q<c<\R d/2<\min(-\alpha_{q-1},\rho)\} \quad ,
\end{equation*}
so that
\begin{equation} \label{eq:final}
\begin{split}
S(d,R) =& \frac{\pi^{d/2}}{\Gamma(d/2)} \mt{\fsub,d/2} +
 \frac{1}{2\pi i} \int_{\Gamma^\prime} ds\ \zeta(s,d)\ R^{2s-d} \mt{\fsub,s} + \\
 &+\sum_{k=0}^{q-1}\  \frac{a_k}{R^{2\alpha_k+d}}\ \zeta(-\alpha_k,d)\quad .
\end{split}
\end{equation}
The continuation of this expression can be easily found by considering those of 
$\mt{\fsub,d/2}$\ in the first term and of $\zeta(s,d)$\ in the second and third ones, 
while the contour is always given by 
$\Gamma'=\{\R s=c; -\alpha_q<c<\min(-\alpha_{q-1},\rho,\R d/2)\}$.

Dimensional poles of this expression may arise from the first and third terms on r.h.s.:
the former is $R$-independent and exactly corresponds to the
integral~\reff{eq:subtracteddefinition} defined in dimensional regularization
with the proper subtraction. Its possible IR divergences are cancelled
by the poles of the last term for $d=-2\alpha_k$, $k=0,\cdots,(q-1)$, which may be 
``physically'' accessible, i.e. be positive, if some $\alpha_k$\ is negative.
The final result is thus IR-finite, as it should be, the series being free from IR 
divergences.
From~\reff{eq:final} it is evident that the UV divergences are the same for the
series and the corresponding integral.
Finally, when $\R d>-2\alpha_0$ the infinite radius limit can be safely extracted 
from~\reff{eq:final}, leading to eq.~\reff{eq:infiniteR} as promised.

\subsection{Evaluating the finite part with a $\vartheta$-function} 
\label{threp}

A general formula for the complex integral in~\reff{eq:final} can be obtained under 
particular assumptions. Performing the appropriate IR subtraction, one always reduces 
oneself to the computation of the integral
\begin{equation}
\label{eq:inttheta}
\frac{1}{2\pi i} \int_\Gamma ds \,\zeta(s,d) R^{2s-d} \mt{f,s} \quad ,
\end{equation}
where, if the function $f$ has the usual asymptotic expansion~\reff{eq:andam} 
and $f(t)\sim t^{-\rho}$\ for $t\to\infty$, then
$\Gamma=\{\R s=c\, ;\, -\alpha_0<c<\min(\rho,\R d/2)\}$.
Let us suppose that
\begin{enumerate}
\item \hspace{0.3cm} $f(s)$ is analytical in the half-plane $\R s>\bar c$, $\bar c\le 0$, 
\item \hspace{0.3cm} $\displaystyle \lim_{\R s\to +\infty} f(s) = 0, 
 \qquad \text{i.e.}\quad \rho>0$,
\item \hspace{0.3cm} $\displaystyle \Imint{>\bar c} f(s)\ e^{ts}  < \infty 
 \qquad \text{for} \quad t>0$.
\end{enumerate}
Under these hypotheses the inversion theorem (see~\cite{Davies}) guarantees that the Laplace 
anti-transform
\begin{equation} \label{eq:laplaceA}
\lit{f,t} =  \Imint{>\bar c} f(s)\ e^{ts} \quad ,
\end{equation}
exists for $t>0$ and 
\begin{equation} \label{eq:lit}
f(s) = \int_0^\infty\!\! dt\, \lit{f,t}\ e^{-st} \qquad \R s>\bar c\quad .
\end{equation}
Using these results, the Mellin transform can be expressed as follows:
\begin{equation} \label{eq:trick}
\mt{f,s} = \int_0^\infty \!\!dt\, t^{s-1} f(t) =
 \int_0^\infty\!\! dt\, t^{s-1} \int_0^\infty \!\!dy\, e^{-ty} \lit{f,y} = 
 \Gamma(s) \int_0^\infty\!\! dy\, \lit{f,y}\, y^{-s} \quad . 
\end{equation}
This relation holds when $-\alpha_0<\R s<\rho$ and it is therefore valid along the contour 
$\Gamma$ of eq.~\reff{eq:inttheta}. 
If $\alpha_0>0$,  $\R s$ can be negative and $f^{(k)}(0)=0$, for 
$k=0,\cdots,\lfloor \alpha_0\rfloor$,
$\lfloor x\rfloor\equiv\max\{n\in{\mathbb N}\, |\, n < x\}$.
Then the simple poles in $s=0,\,-1,\,-2,\cdots,\,-\lfloor \alpha_0\rfloor$ of $\Gamma(s)$ 
in~\reff{eq:trick} 
are cancelled by corresponding first order zeros of the integral of the Laplace 
anti-transform
\begin{equation*}
\int_0^\infty \!\!dy\, \lit{f,y}\,y^{k} = (-1)^k\,f^{(k)}(0) = 0 
 \quad \text{for}\quad k=0,\,1,\,\cdots,\,\lfloor \alpha_0\rfloor \quad\text{if}
 \quad \alpha_0 > 0 \quad ,
\end{equation*}
resulting in a Mellin transform well defined in the whole interval $-\alpha_0<\R s<\rho$. 
Using~\reff{eq:trick} and the analytic extension of $\zeta(s,d)$ in terms of the 
$\vartheta$-function (see Appendix~\ref{Appzeta})
\begin{equation*}
\zeta(s,d)=\frac{\pi^s}{\Gamma(s)}\left\{\frac{1}{s-d/2} - \frac{1}{s} +
 \int_1^{\infty}\!\!dy\, \left(y^{s-1}+y^{d/2-s-1}\right)\,
 \left[\vartheta^d(y) - 1 \right] \right\} \quad ,
\end{equation*}
it is not difficult to compute the complex integral in~\reff{eq:inttheta}
by applying standard residue techniques. The result is:
\begin{equation} \label{eq:thetarepres}
\begin{split}
\frac{1}{2\pi i} \int_\Gamma ds \, \zeta(&s,d) R^{2s-d} \mt{f,s} =\\
= &-\frac{\pi}{R^{d-2}} \int_1^\infty \!\! dy\ \left( \lit{f,y \pi R^2}\, y^{-d/2} +
 \lit{f,\pi R^2/y}\, y^{-2} \right) +\\
 &+\frac{\pi}{R^{d-2}} \int_1^\infty \!\! dy\ [\vartheta^d(y) -1]
 \left( \lit{f,\pi R^2 y} + y^{d/2-2} \lit{f,\pi R^2/y} \right) \quad .
\end{split}
\end{equation}
The two integrals converge under the assumed hypotheses~\footnote{This is easily understood 
from the asymptotic behaviour of the Laplace anti-transform (see~\cite{Davies}):
\begin{equation*}
\lit{f,t}\sim \begin{cases} t^{\rho-1} & t\to 0 \\  t^{-(\alpha_0+1)} & t\to \infty 
 \end{cases}, \qquad  \text{which follows if} \qquad
f(t)\sim \begin{cases} t^{-\rho} & t\to\infty \\ t^{\alpha_0} & t\to 0 \end{cases} \quad .
\end{equation*} 
}.
Equation~\reff{eq:thetarepres} is a useful representation for the complex integral 
in~\reff{eq:final}, which allows an expansion of the result in terms of the parameters 
in $f$ or even to perform a numerical integration.
The results of ref.~\cite{GravitonLoops} are easily obtained as particular applications 
of~\reff{eq:thetarepres}.
It is worth noting that one could have derived the same result by giving a definition
of the series directly in terms of the $\vartheta$-function, never using $\zeta$-functions,
and this is exactly the procedure adopted in ref.~\cite{GravitonLoops}.
Of course, to define the series using $\vartheta$-functions one has to assume the same
hypotheses we imposed on the function $f$ to guarantee the existence of the 
Laplace anti-transform. In this respect, our definition seems more general and
more suitable to isolate the divergent from the finite term.

\section{A sample computation: the Casimir energy}
\label{example}

To give an explicit example of our procedure, we compute the Casimir energy $\mathcal{E}(R)$
for a scalar field with mass $m$ and periodicity conditions in a space geometry 
$\mathbb{T}^{\bar d}\times{\mathbb R}^{\bar D}$, corresponding to $\bar d$ compactified 
dimensions with radius $R$.
This is defined as the $R$-dependent part of the zero point energy $V(R)$
of the $(\bar D+\bar d)$-dimensional theory multiplied by the volume 
$(2\pi R)^{\bar d}$ of the compact space:
\begin{equation} \label{eq:casimir}
\mathcal{E}(R) = (2\pi R)^{\bar d}\ \big[ V(R) - V(\infty) \big] \quad ,
\end{equation}
Continuing the physical dimensions $\bar D$, $\bar d$ to generic values
\begin{equation} 
D=\bar D - \epsilon \, ; \qquad d = \bar d - \eta \quad ,
\end{equation}
the one-loop contribution to the zero point energy is given by
\begin{equation} \label{eq:ZPenergy}
V(R) = \frac{1}{2} \, \frac{\mu^{\epsilon+\eta}}{(2\pi R)^d} \sum_{n\in\mathbb{Z}^d} 
 \int \frac{d^Dp}{(2\pi)^D} \log\left(p^2+n^2/R^2+m^2\right) \quad .
\end{equation}
To extract $\mathcal{E}(R)$ we first compute $V(R)$, then extract the 
limit $\epsilon\to 0$, $\eta\to 0$ and finally multiply by the volume factor.

For convenience let us call $V_0(R)$ the zero mode contribution in~\reff{eq:ZPenergy}.
We will show explicitly that performing first the series and then the integral or vice versa
leads to the same result (see property 2 in Appendix~\ref{properties}) and there is 
no ambiguity. Let us perform the integral first.
The function $f(x)=\log(x+c)$ has the asymptotic behaviour $f(x)\sim \log x$ for
$x\to\infty$ and $f(x)\sim$~const. for $x\to 0$; this means that the dimensionally
regulated integral needs subtractions to be defined. Performing one subtraction, i.e.
considering $f_{\text{sub}}(x)=\log(x/c+1)$, we can fix $-2<D<-1$, obtaining
\begin{equation}
V(R) - V_0(R) = - \frac{1}{2} \, \mu^{\epsilon+\eta}\, \frac{\pi^{D/2}}{(2\pi)^D} \Gamma(-D/2)
 \, \frac{1}{(2\pi R)^d} \sumprime{n}{d} (n^2/R^2+m^2)^{D/2} \quad .
\end{equation}
Because $D$ is negative and the function $g(x)=(x+m^2)^{D/2}$ has the asymptotic behaviour
$g(x)\sim x^{D/2}$ for $x\to\infty$ and $g(x)\sim$~const. for $x\to 0$, the series
is defined by~\reff{eq:subdef} with 
$\Gamma=\{\R s=c; 0<d/2<c<-D/2<1 \}$  and without any subtraction:
\begin{equation}
V(R) - V_0(R) = - \frac{1}{2} \, \mu^{\epsilon+\eta} \, \frac{\pi^{D/2}}{(2\pi)^D} \Gamma(-D/2)
 \, \frac{1}{(2\pi R)^d} \frac{1}{2\pi i} \int_\Gamma ds\ \zeta(s,d)\ R^{2s} \mt{g,s} \quad .
\end{equation}
Even if the Mellin transform is easy to derive
\begin{equation*}
\mt{g,s} = \frac{\Gamma(s)\Gamma(-D/2-s)}{\Gamma(-D/2)} \left(m^2\right)^{D/2+s} \quad ,
\end{equation*}
the complex integral along the contour $\Gamma$ cannot be solved with simple 
residue techniques because of the non-trivial behaviour of the integrand at infinity.
However, $g(s)$ is analytic in the half-plane $\R s>-m^2$ and the requirements
needed to apply the representation in terms of the $\vartheta$-function are fulfilled.
From eqs.~\reff{eq:final} and \reff{eq:thetarepres} with a Laplace anti-transform 
\begin{equation*}
\lit{g,y} = \frac{e^{-m^2 y}}{\Gamma(-D/2)} y^{-D/2-1} \quad ,
\end{equation*}
we obtain
\begin{equation}  \label{eq:casintermediate}
\begin{split}
V(R) - V_0(R) = -\frac{1}{2} & \mu^{\epsilon+\eta}\ \frac{\pi^{(D+d)/2}}{(2\pi)^{D+d}}
 \Gamma\left(-\frac{D+d}{2}\right) \left(m^2\right)^{(D+d)/2} + \\
 \frac{1}{2} \frac{\mu^{\epsilon+\eta}}{(2\pi R)^{D+d}} \bigg\{
  & \int_1^\infty\!\! dy\ \left( e^{-\pi y (mR)^2} y^{-(D+d)/2-1} +
  e^{-\pi (mR)^2/y} y^{D/2-1} \right) - \\
 & \int_1^\infty\!\! dy\ [\vartheta^d(y)-1] \left(e^{-\pi y (mR)^2} y^{-D/2-1} +
   e^{-\pi (mR)^2/y} y^{(D+d)/2-1} \right) \bigg\} \quad .
\end{split}
\end{equation}
We recognize the infinite radius contribution in the first $R$-independent term,
while the remaining terms are finite.
More exactly, the integral in the second line is convergent when
$\R D<0$ as supposed from the beginning to regularize the dimensional integral.
Therefore, when $\R D$ is increased to (physical) values $\R D>0$ a 
divergence appears which, however, depends only on $D$ but not on the value 
of $d$. This means that in some way this must be a ``zero mode''
divergence and this becomes evident by rewriting
\begin{equation} \label{eq:zeromode}
\int_1^\infty \!\! dy \, e^{-\pi (mR)^2/y} y^{D/2-1} =
 \left[\pi (mR)^2\right]^{D/2} \Gamma(-D/2) -
 \int_1^\infty \!\! dy \, e^{-\pi y (mR)^2} y^{-D/2-1} \quad .
\end{equation}
The first term on the r.h.s. of~\reff{eq:zeromode} exactly cancels the zero mode
contribution $V_0(R)$ in~\reff{eq:casintermediate} and we get the final result
\begin{equation} \label{eq:casimirfinal}
\begin{split}
V(R) = -\frac{1}{2} & \mu^{\epsilon+\eta}\ \frac{\pi^{(D+d)/2}}{(2\pi)^{D+d}}
 \Gamma\left(-\frac{D+d}{2}\right) \left(m^2\right)^{(D+d)/2} + 
 \frac{1}{2} \frac{\mu^{\epsilon+\eta}}{(2\pi R)^{D+d}} \times \\ \times  \bigg\{
 & \int_1^\infty \!\! dy\ e^{-\pi y (mR)^2} \left( y^{-(D+d)/2-1} +
   y^{-D/2-1} \right) - \\
 & \int_1^\infty \!\! dy\ [\vartheta^d(y)-1] \left(e^{-\pi y (mR)^2} y^{-D/2-1} +
   e^{-\pi (mR)^2/y} y^{(D+d)/2-1} \right) \bigg\} \quad .
\end{split}
\end{equation}
The first term is the ordinary divergent renormalization of the 
cosmological constant, which can be put to zero with a suitable counterterm
if we accept the usual fine tuning.
Whatever scheme of renormalization one chooses, the radius dependent part $V(R)-V(\infty)$
of the zero point energy is non ambiguous and finite.
We can therefore safely extract the limit $\epsilon\to 0$, $\eta\to 0$ and insert the result
for $V(R)-V(\infty)$ in eq.~\reff{eq:casimir} to extract the Casimir energy
\begin{equation}
\begin{split}
\mathcal{E}(R) =& \ \frac{1}{2}\ \frac{1}{(2\pi R)^{\bar D}}\ \bigg\{
 \int_1^\infty \!\! dy\ e^{-\pi y (mR)^2} \left( y^{-(\bar D+\bar d)/2-1} +
   y^{-\bar D/2-1} \right) - \\
 & \int_1^\infty \!\! dy\ [\vartheta^{\bar d}(y)-1] \left(e^{-\pi y (mR)^2} y^{-\bar D/2-1} +
   e^{-\pi (mR)^2/y} y^{(\bar D+\bar d)/2-1} \right) \bigg\} \quad .
\end{split}
\end{equation}

The same result can be obtained by performing first the series.
Again, subtractions are necessary if one does not introduce an IR cut-off 
in the Mellin transform. The function $f(x)=\log(x+p^2+m^2)$ has an expansion around
$x=0$ as in~\reff{eq:andam} with $\alpha_0=0$, $\alpha_1=1$; we can therefore apply
definition~\reff{eq:subdef} with one subtraction and, choosing the contour
$\Gamma = \{\R s=c;-1<d/2<c<-1/2 \}$, we obtain:
\begin{equation} \label{eq:seriesfirst}
\begin{split}
V(R)-V_0(R) =& \frac{1}{2}\ \frac{\mu^{\epsilon+\eta}}{(2\pi R)^d} \int \frac{d^Dp}{(2\pi)^D}\
 \zeta(0,d) \log\left(p^2+m^2\right) + \\ &+\frac{1}{2} \frac{\mu^{\epsilon+\eta}}{(2\pi R)^d}
 \int \frac{d^Dp}{(2\pi)^D} \frac{1}{2\pi i} \int_\Gamma ds\ \zeta(s,d)\ R^{2s} 
 \mt{f_{\text{sub}},s} \quad ,
\end{split}
\end{equation}
with
\begin{equation*}
\mt{f_{\text{sub}},s} = -\left[p^2+m^2\right]^s \Gamma(s) \Gamma(-s) \quad .
\end{equation*}
The complex integral along the contour $\Gamma$ cannot be solved with simple residue
techniques nor we can apply the $\vartheta$-function representation ($f_{\text{sub}}$
diverges at infinity), but we can make use of a standard trick. Namely, defining
\begin{equation*}
S(p^2+m^2) = \frac{1}{2\pi i} \int_\Gamma ds\ \zeta(s,d)\ R^{2s} \mt{f_{\text{sub}},s}
\end{equation*}
we know how to sum the series $d S(p^2+m^2)/dp^2$ because 
$df_{\text{sub}}/dp^2=1/(x+p^2+m^2)$
goes to zero for $x\to\infty$. Then we can deduce the expression of $S(p^2+m^2)$
except for an unknown function independent of $p^2$, which however is irrelevant as its 
dimensional integral in $p$ gives zero. In particular we have:
\begin{equation} \label{eq:derivseries}
\begin{split}
\frac{d}{dp^2} S(p^2+m^2) =& \frac{1}{2\pi i} \int_\Gamma ds\ \zeta(s,d)\ R^{2s}
 \frac{d}{dp^2} \mt{f_{\text{sub}},s} \\ 
 =& \frac{1}{2\pi i} \int_{\Gamma^\prime} ds\ \zeta(s,d)\ R^{2s}
 \frac{d}{dp^2} \mt{f_{\text{sub}},s} - \zeta(0,d) \frac{1}{p^2+m^2} \quad ,
\end{split}
\end{equation}
where the contour $\Gamma$ has been moved into a new one,
$\Gamma^\prime=\{\R s=c; d/2<0<c<1 \}$ and a residue contribution in $s=0$ has 
been isolated.
The first term corresponds to the (dimensionally regularized) series of the function
$d\log(p^2+n^2/R^2+m^2)/dp^2$, defined without subtractions by choosing the contour
$\Gamma^\prime$, and it can be easily computed using the 
$\vartheta$-function representation \reff{eq:thetarepres}. 
We do not show the details of this calculation
but we only notice that using~\reff{eq:derivseries} to extract $S(p^2+m^2)$ and finally
plugging the result into~\reff{eq:seriesfirst}, 
the residue term in the former equation cancels the subtraction term of the latter
and we end up with
\begin{equation}
\begin{split}
V(R)-V_0(R) = & \frac{1}{2}\ \mu^{\epsilon+\eta} \int \frac{d^Dp}{(2\pi)^D} \,
 \bigg\{ -\frac{\pi^{D/2}}{(2\pi)^d} \Gamma(-D/2) \left[p^2+m^2\right]^{d/2} + \\ 
 \frac{1}{(2\pi R)^d} 
 \bigg[ &\int_1^\infty \!\! dy\ \left( e^{-\pi y(m^2+p^2)R^2} y^{-d/2-1}
 + e^{-\pi (m^2+p^2)R^2/y} y^{-1} \right) + \\
 & \int_1^\infty \!\! dy\ [\vartheta^d(y)-1] \left( e^{-\pi y(m^2+p^2)R^2} y^{-1} +
  e^{-\pi (m^2+p^2)R^2/y} y^{d/2-1} \right) \bigg] \bigg\} \quad .
\end{split}
\end{equation}
The result of the dimensional integration over $p$ coincides
with eq.~\reff{eq:casintermediate}, from which the final 
expression~\reff{eq:casimirfinal} follows.

\section{Conclusions}
\label{conclusions}

We discussed in detail a general procedure to dimensionally regularize divergent series.
The novelty of the method consists in using a suitable combination of two well-known tools, 
i.e. Mellin transform and analytic extension of special functions,
to provide a continuation of the series in the number of dimensions.

The virtue of conventional dimensional regularization is to preserve all
the symmetries of the action that do not depend on the dimensionality, in particular
gauge invariance and supersymmetry (if the dimensional reduction scheme is used).
The same happens with dimensional regularization of sums, making this technique
a natural choice to handle divergences of field theory.
In this respect, our analysis should contribute to clarify some controversial aspects
about the computation of quantum corrections in supersymmetric theories with extra dimensions
\cite{BHN,NG,Quiros,CP}. 
In particular, we have shown that there is no ambiguity in exchanging 
the series and the integral over loop momenta if both are consistently regularized with 
dimensional regularization (see Property 2 in Appendix~\ref{properties} and the example in 
Sec.~\ref{example}). The same is not true for instance if the sum over Kaluza--Klein modes is 
truncated by a raw cut-off. 

Our definition of sums in complex dimensions by using a generalized zeta function
is particularly suited to isolate the divergence from the finite part and applies
to a large class of functions. Moreover, it leads to simple computations and it is
valid for an arbitrary number of physical compact dimensions.
The idea of dimensionally regularized series was applied in ref.~\cite{GravitonLoops}
to compute a number of loop sums by using a representation in terms of $\vartheta$-functions.
We obtain the results of ref.~\cite{GravitonLoops} as particular cases of our
general formulae.
A different method was proposed in~\cite{Nibbelink},
which however applies only to the case of one physical compact dimension.
Our procedure has no such limitation and it is therefore more general; 
in the case of only one physical compact dimension it gives the same result 
as~\cite{Nibbelink}, as we have explicitly checked for particular examples.

Although the class of functions we considered is not the most general one, the 
basic idea may be applied to more complicated cases by using suitable zeta 
functions. In particular, our formulae are specific to series that appear
in theories with toroidal compact dimensions and scalar fields with simple periodical
conditions. In the case of twisted periodic conditions or theories with fermions,
our method should be easily extended introducing the following zeta function
\begin{equation*}
\zeta(s,d\,|\,a) \equiv \sumprime{n}{d}\ \frac{1}{[(n+a)^2]^s} \quad , 
 \quad -1/2<a\le 1/2 \quad ,
\end{equation*}
whose properties are sketched in Appendix~\ref{Appzeta}.
The way to treat spinors in the dimensional continuation is similar as in conventional DR,
see~\cite{Collins,Nibbelink}.

Even when the compact manifold is not toroidal~(see for example \cite{CW}) 
or the single dimensions have different radii, we see no obstacle in principle to
apply the approach of dimensional regularization, maybe taking advantage of
more general special functions such as those introduced in~\cite{zfunction}.
In these cases, however, computations may become involved and particular recipes
specific to the case may be simpler to use.

\section*{Acknowledgements}

We thank S. Caracciolo, A. Pelissetto and R. Rattazzi for a critical reading of 
the manuscript. R.C. acknowledges in a special way R. Rattazzi for many useful discussions and 
suggestions. Particular thanks go to A. Strumia for enlightening us on the 
importance and usefulness of {\it Remnant Functions}.
This work is partially supported by the EC under TMR contract HPRN-CT-2000-00148.
\vspace{-0.1cm}


\appendix
\numberwithin{equation}{section}

\section{Mellin transform}
\label{appmellin}

Let us briefly recall the definition of the Mellin transform~\cite{Davies, GraRyz}.
Given a function $f$, if
$\alpha, \beta\in{\mathbb R}$, $\alpha<\beta$ exist such that
\begin{equation*}
\int_0^{\infty}\!\!dt\ |f(t)|\ t^{\rho-1}\  < \ \infty\ ;\  
\ \ \forall \rho\ :\ \alpha<\rho<\beta \quad ,
\end{equation*}
then one can define the Mellin transform of $f(t)$,
\begin{equation}
\mt{f,s}= \int_0^{\infty}\!\!dt\ f(t)\ t^{s-1} \quad ,
\end{equation}
which is an analytic function of $s\in{\mathbb C}$ in the strip 
$\alpha<\R s<\beta$. This inversion formula holds
\begin{equation}
f(t) = \frac{1}{2\pi\,i} \int_{c-i\infty}^{c+i\infty}\!\!ds\, 
 \mt{f,s} \,t^{-s}\ \ ; \ \ \alpha<c<\beta \quad .
\label{AppMellin1}
\end{equation}
Obviously it is a linear integral transformation. 
It has very useful applications and remarkable properties, 
see, for instance,~\cite{Davies}.

\section{Properties of $\zeta(s,d)$}
\label{Appzeta}

We define, for integer $d$
\begin{equation}
\zeta(s,d) \equiv \sumprime{n}{d}\ \frac{1}{(n^2)^s} \quad ,
\end{equation}
which is absolutely convergent for $\R s>d/2$. 
This function is a particular case of the more general zeta functions 
(see, for example,~\cite{zfunction} and \cite{luscher}). 
Let us write an explicit expression for the analytic continuation of $\zeta(s,d)$\ 
to the whole complex plane in $s$\ and, eventually, extend its definition also 
for all complex $d$. The function 
(related to Jacobi's $\vartheta_3$~\cite{GraRyz}),
\begin{equation}
\vartheta(t) \equiv \sum_{n=-\infty}^{+\infty} \, e^{-\pi\,t\,n^2} \quad ,
\end{equation}
has the following modular property, which is easily derived from 
Poisson's resummation formula~\cite{Davies}:
\begin{equation}
\vartheta(t) = \frac{1}{t^{1/2}}\, \vartheta(1/t) \quad .
\label{proptheta}
\end{equation}
By means of
\begin{equation}
\frac{1}{(n^2)^s} = \frac{1}{\Gamma(s)}\int_0^{\infty}\!\!dt\ 
 t^{s-1}\ e^{-t\,n^2} \quad ,
\label{schwinger}
\end{equation}
we may write
\begin{equation}
\zeta(s,d)=
\frac{\pi^s}{\Gamma(s)}\int_0^{\infty}\!\!dt\ 
 t^{s-1}\ \left[\vartheta^d(t) - 1 \right] \quad ;
\label{zetarepr1}
\end{equation}
incidentally, this tells us that the Mellin transform of 
$[\vartheta^d(t) - 1]$\ is $\pi^{-s}\,\Gamma(s)\,\zeta(s,d)$.

For $\R s>0$, $\R s>d/2$, 
it follows from~\reff{proptheta} that
\begin{equation*}
\int_0^1\!\!dt\, t^{s-1}\,\left[\vartheta^d(t) - 1 \right] =
 \int_1^{\infty}\!\!dt\, t^{d/2-s-1}\,\left[\vartheta^d(t) - 1 \right] +
 \frac{1}{s-d/2} - \frac{1}{s} \quad ,
\end{equation*}
which gives
\begin{equation}
\zeta(s,d)=\frac{\pi^s}{\Gamma(s)}\left\{\frac{1}{s-d/2} - \frac{1}{s} + 
 \int_1^{\infty}\!\!dt\, \left(t^{s-1}+t^{d/2-s-1}\right)\,
 \left[\vartheta^d(t) - 1 \right] \right\} \quad .
\label{zetaestesa}
\end{equation}
This expression represents the analytic continuation of $\zeta(s,d)$\ 
in both $s$\ and $d$. It is easy to demonstrate the following properties:
\begin{enumerate}
\item $\zeta(s,d)$\ is a meromorphic function, having
a simple pole for $s = d/2$, $d\neq 0$, with residue
 \begin{equation*}
  \text{Res} \, \big\{ \zeta(s,d);s=d/2\big\} = \frac{\pi^{d/2}}{\Gamma(d/2)} \quad ;
 \end{equation*}
\item $\zeta(s,0) = 0$, $\forall s \in {\mathbb C}$;
\item $\zeta(0, d) = -1$, for $d\neq 0$;
\item $\zeta(-n, -2n) = (-1)^n\,n!/\pi^n$, for $n\in{\mathbb N}$, $n\neq 0$;
\item $\zeta(-n,d) = 0$, for $d\neq -2n$ and $n\in{\mathbb N}$, $n\neq 0$;
\item Given the symmetry $s \rightarrow d/2-s$\ of the terms inside the brackets in
\reff{zetaestesa}, we have
 \begin{equation*}
  \frac{\Gamma(s)\,\zeta(s,d)}{\pi^{s}} = 
   \frac{\Gamma(d/2-s)\,\zeta(d/2-s,d)}{\pi^{d/2-s}} \quad .
 \end{equation*}
This is the so-called {\it reflection formula}, which is also valid in the 
more general case of Epstein zeta functions~\cite{zfunction}.
\end{enumerate}
We now derive a ``convolution'' property of $\zeta(s,d)$, used in 
Appendix~\ref{AppSS}, eq.~\reff{eq:propSS}.
For $p,q>0$ and $\R s > (p+q)/2$,
\begin{equation}
\begin{split}
\zeta(s,p+q) &= \sumprime{m}{p}\sumprime{n}{q}\frac{1}{(m^2+n^2)^s} +
 \sumprime{m}{p}\frac{1}{(m^2)^s} +\sumprime{n}{q}\frac{1}{(n^2)^s} \\
 &= \sumprime{m}{p}\sumprime{n}{q}\frac{1}{(m^2+n^2)^s} +\zeta(s,p)+\zeta(s,q)\quad .
\label{relsomme}
\end{split}
\end{equation}

We can write the argument of the double sum using~\reff{schwinger} and then introduce 
the Mellin representation for one of the two exponential factors, getting the result
\begin{equation}
\begin{split}
\sumprime{m}{p}\sumprime{n}{q}\frac{1}{(m^2+n^2)^s} &=  
 \frac{1}{\Gamma(s)}\int_0^{\infty}\!\!dt\ t^{s-1}\,\sumprime{n}{q} e^{-t\,n^2}\, 
 \sumprime{m}{p} \frac{1}{2\pi i}  \int_{c-i\infty, c>0}^{c+i\infty}\!\!\!dw\ \,
 \Gamma(w)\;(m^2 \,t)^{-w}  \\
 &=\frac{1}{\Gamma(s)}\frac{1}{2\pi\, i}\,\int_{c-i\infty}^{c+i\infty}\!\!\!dw\ 
 \Gamma(w)\,\zeta(w,p)\,\Gamma(s-w)\,\zeta(s-w,q) \quad ,
\label{proprdimtran1}
\end{split}
\end{equation}
where, in the last line, $p/2<c<\R s -q/2$. 
Taking into account relation~\reff{relsomme}, we have~\footnote{Note that it is not 
possible to close the contour of integration at 
infinity, given the asymptotic behaviour of the integrand.}
\begin{equation}
\begin{split}
\frac{1}{2\pi\, i}\,\int_{c-i\infty}^{c+i\infty}\!\!\!dw\ \Gamma(w)\,\zeta(w,p)\,
 \Gamma(s-w)\,\zeta(s-w,q) &=\\
 \Gamma(s)\,\zeta(s,p+q)&\ - \Gamma(s)\,\zeta(s,p)- \Gamma(s)\,\zeta(s,q) \quad .
\end{split}
\label{proprdimtran2}
\end{equation}
This relation is still valid if one considers the analytic extensions of 
the functions involved.

Next, we determine the asymptotic behaviour of $\zeta(s,d)$\ for large $\R s$\ and $d>0$. 
Let us observe that we may write
\begin{equation}
\vartheta^d(t) = 1 + 2d\, e^{-\pi\, t} + \sum_{k=0}^{\infty}\, 
 {\mathcal N}_{k}(d)\, e^{-\alpha_k\,\pi\,t} \quad ,
\label{serietheta}
\end{equation}
where ${\mathcal N}_{k}(d)$ are real coefficients and $4\le\alpha_0<\alpha_1<\cdots$. 
For large $k$, the series is asymptotic to the one with $\alpha_k = k^2$\ and 
${\mathcal N}_{k}(d)$ is the number of points in ${\mathbb Z}^d$\ with a distance from 
the origin bounded between $k$\ and $(k+1)$. So ${\mathcal N}_{k}(d)\sim\Omega_d\,k^{d-1}$
(for $k\gg 1$). For $\R s > d/2$\ we may insert~\reff{serietheta}\ into~\reff{zetarepr1}, 
obtaining
\begin{equation*}
\zeta(s,d) = 2d + \sum_{k=0}^{\infty}\, {\mathcal N}_{k}(d)\, \alpha_k^{-s} \quad ,
\end{equation*}
so that
\begin{equation}
|\,\zeta(s,d)-2d\,| \le \alpha_0^{-\R s}\, \sum_{k=0}^{\infty}\, 
 {\mathcal N}_{k}(d)\, (\alpha_k/\alpha_0)^{-\R s} \quad .
\end{equation}
The series on the r.h.s.\ is always convergent for $\R s > d/2$, 
and therefore $\zeta(s,d)- 2d \sim \alpha_0^{-\R s}$\ for $\R s \to +\infty$. 
Using the reflection formula we immediately obtain the asymptotic behaviour for 
$\R s \to -\infty$: in this case the function is 
unbounded~\footnote{We use the fact that for $x\in \mathbb{R}^+$,
$m\equiv\lfloor x\rfloor$, $q\equiv x-\lfloor x\rfloor$, one has 
 \begin{gather*}
  \Gamma(- x) = (-1)^m \frac{\Gamma(q+1)\,\Gamma(-q)}{\Gamma(x+1)} \quad , \\
  \Gamma(-x)= (-1)^m \Gamma(q+1)\,\Gamma(-q) \frac{x^{-x}\,e^{x}}{\sqrt{2\pi x}}
   (1 + O(x^{-1})) \quad .
 \end{gather*}
 }:
\begin{equation}
\zeta(-x,d) = (-1)^{\lfloor x\rfloor} \frac{2d(x/\pi)^{2x+d/2}\,
 e^{-2x}}{\Gamma( x-\lfloor x\rfloor+1)\,\Gamma(\lfloor x\rfloor-x)}
 (1+O(x^{-1})) \quad , \quad x\in \mathbb{R}^+ \quad .
\end{equation}

\subsection*{A useful generalization}

In some cases of interest (such as in theories with fermions) it is useful to introduce 
the following generalization of our zeta function. 
Given a constant $-1/2< a\le 1/2$, we define
\begin{equation}
\label{eq:zetagen}
\zeta(s,d\,|\,a) \equiv \sumprime{n}{d}\ \frac{1}{[(n+a)^2]^s} \quad ,
\end{equation}
where now $\sum'$\ means that for $a=0$\ we omit the term $n=0$ in the sum. 
This series converges for $\R s > d/2$. Obviously $\zeta(s,d\,|\,0)=\zeta(s,d)$. 
To get the analytic continuation of this function one proceeds
as in the case of $\zeta(s,d)$. Introducing the $\vartheta$-function
\footnote{It has the asymptotic behaviour:
 \begin{equation}
  \vartheta (t|a) \sim e^{-\pi ta^2} \quad \text{for} \quad t\to\infty \; ; \qquad
  \vartheta (t|a) \sim t^{-1/2} \quad \text{for} \quad t\to 0 \quad .
 \end{equation} 
}
\begin{equation}
\label{eq:thetagen}
\vartheta(t\,|\,a) \equiv \sum_{n=-\infty}^{+\infty} \, e^{-\pi\,t\,(n+a)^2} \quad ,
\end{equation}
from the Poisson resummation formula easily follows the modular property
\begin{equation}
\vartheta(t|a) = \frac{e^{-\pi\,a^2\,t}}{t^{1/2}}\vartheta(1/t\,|\,i\,a\,t) \quad ,
\end{equation}
The analogues of eqs.~\reff{zetarepr1} and \reff{zetaestesa} are
\begin{gather}
\label{zetagenrepr1} 
\zeta(s,d\,|\,a)=
\frac{\pi^s}{\Gamma(s)}\int_0^{\infty}\!\!dt\, t^{s-1}\,
 \left[\vartheta^d(t\,|\,a) - \delta_{a,0} \right] \\[0.3cm]
\begin{split}
\zeta(s,d\,|\,a) = \frac{\pi^s}{\Gamma(s)} \bigg\{  &\frac{1}{s-d/2} - 
 \frac{1}{s}\,\delta_{a,0}+
 \int_1^{\infty}\!\!dt\, t^{s-1}\left[\vartheta^d(t\,|\,a) - \delta_{a,0} \right] + \\
 & \int_1^{\infty}\!\!dt\, t^{d/2-s-1}\,
 \left[e^{-\pi\,a^2\,d/t}\vartheta^d(t\,|\,ia/t) - 1 \right] \bigg\} \quad ,
\end{split}
\label{zetagenestesa}
\end{gather}
where the last expression has a meromorphic extension with the same general 
properties as those of $\zeta(s,d)$.
We remark that the reflection formula does not hold in the general case $a\neq 0$.

\section{Properties of dimensional continuation of sums}
\label{properties}

We present here some basic properties of the dimensionally regularized series
as defined by~\reff{repres}.
They are the analogues of those discussed in~\cite{Collins},
valid for dimensionally regularized integrals.

\vspace{0.5cm}
\noindent \textit{Property} 1.
\label{AppSS}
\begin{equation}
 \sum_{n\in{\mathbb{Z}}^p}\, \sum_{m\in{\mathbb{Z}}^q}\,f(n^2+m^2) =
 \sum_{n\in{\mathbb{Z}}^{p+q}}\, f(n^2) \quad .
\end{equation}

\vspace{0.5cm}
\textit{Proof}. \hspace{0.05cm}  
Consider a Mellin-transformable function $f(x)$, assuming for simplicity
that $f(x)\to 0$ for $x\to 0$ and $f(x)\sim x^{-\rho}$, $\rho>0$ for $x\to\infty$,
such that its series can be defined without IR subtractions, which are irrelevant
to this discussion.
Let us denote by $\mt{f,s;y}$ the Mellin transform of $f(x+y)$ with respect to
the variable $x$:
\begin{equation*}
\mt{f,s;y} \equiv \int_0^{\infty} \!\!dx \, f(x+y)\, x^{s-1} \quad .
\end{equation*}
Transforming also on $y$ and performing a change of variables we get
\begin{equation}
\begin{split}
\mt{f,s;w} & \equiv \int_0^{\infty} \!\!dy \, \mt{f,s;y} \, y^{w-1} = 
 \int_0^{\infty}\!\!dt\,f(t)\, t^{s+w-1}\,\int_0^1\!\! dv\, v^{s-1}(1-v)^{w-1}  \\
 &= \frac{\Gamma(s)\Gamma(w)}{\Gamma(s+w)}\, \mt{f,s+w} \quad ,
\end{split}
\label{eq:doubleM}
\end{equation}
with $\R s, \R w >0$, $\R s + \R w <\rho$ for these expressions to make sense.
Applying the definition of regularized sums and using~\reff{eq:doubleM}, it follows that
\begin{equation*}
\sumprime{m}{p}\sumprime{n}{q}\, f(n^2+m^2) =
 \frac{1}{(2\pi\, i)^2} \int_{\Gamma_w}\!\!\!dw\ \int_{\Gamma_s}\!\!\!ds\ \zeta(w,p)
 \zeta(s,q)\frac{\Gamma(s)\Gamma(w)}{\Gamma(s+w)}\ \mt{f,s+w} \quad .
\end{equation*}
By making a change of variable in the double complex integral and taking into account 
eqs.~\reff{proprdimtran1}\ and~\reff{proprdimtran2}, we get
\begin{equation}
\begin{split}
\sumprime{m}{p}\sumprime{n}{q}\, f(n^2+m^2) &=
 \frac{1}{(2\pi\, i)^2} \int_{\Gamma_u}\!\!\!du\, \frac{\mt{f,u}}{\Gamma(u)}\ 
 \int_{\Gamma_w}\!\!\!dw\ \Gamma(w)\zeta(w,p)\Gamma(u-w)\zeta(u-w,q) \\
 &=\frac{1}{2\pi\, i} \int_{\Gamma_u}\!\!\!du \,\mt{f,u}\, 
 \left[-\zeta(u,p) -\zeta(u,q) + \zeta(u,p+q)\right]  \\
 &= -\sumprime{m}{p}\,f(m^2) \,-\sumprime{n}{q}\,f(n^2) + \sumprime{n}{p+q} f(n^2) \quad ,
\end{split}
\label{eq:propSS}
\end{equation}
where $\Gamma_u$\ is a contour laying in the half-plane $\R w<\rho$.
This is exactly the equality we are looking for.

\vspace{0.5cm}
\noindent \textit{Property} 2.
\begin{equation} \label{eq:P2}
\int d^Dp \sumprime{n}{d}\ f(p^2,n^2) = 
 \sumprime{n}{d} \int d^Dp\ f(p^2,n^2) \quad .
\end{equation}

\vspace{0.5cm}
\textit{Proof}. \hspace{0.05cm}  
It is convenient to introduce an auxiliary function
\begin{equation*}
f(a,p^2,n^2)=f(p^2,n^2)\ e^{-a(p^2+n^2)} \quad ,
\end{equation*}
for which the right- and left-hand sides of~\reff{eq:P2} become
\begin{subequations} \label{eq:side}
\begin{align}
\text{l.h.s. :} & \qquad \frac{\Omega_D}{2} \int_0^\infty \!\! dx\ x^{D/2-1}\
 \frac{1}{2\pi i} \int_\Gamma ds\ \zeta(s,d) \;  
 \int_0^\infty \!\! dy\ y^{s-1}\ f(a,x,y)  \tag{\ref{eq:side}L} \\ 
\text{r.h.s. :} & \qquad \frac{\Omega_D}{2}\ \frac{1}{2\pi i} 
 \int_\Gamma ds\ \zeta(s,d) \; \int_0^\infty \!\! dy\ y^{s-1}\
 \int_0^\infty \!\! dx\ x^{D/2-1}\ f(a,x,y)  \quad .\tag{\ref{eq:side}R} 
\end{align}
\end{subequations}
Thanks to the exponential factor, the series and the integral in both expressions
have been defined without recourse to subtractions, 
by simply choosing a large enough value for 
(the real part of) $D$ and $d$  to have IR convergence.
Also, the contour $\Gamma=\{\R s=c\, ; c>d/2 \}$ can be defined to be the same 
in~(\ref{eq:side}L) and (\ref{eq:side}R) by fixing $c$ sufficiently large.
Then, property~2 for the auxiliary function follows trivially from exchanging
the integrals over $x$ and $y$ and (\ref{eq:side}L), (\ref{eq:side}R) are both equal 
to the same function $I(D,d,a)$ analytic in its variable.
The analytical continuation of $I(D,d,a)$ down to smaller
$D$ and $d$ is still given by (\ref{eq:side}L), (\ref{eq:side}R), but now with 
subtractions made.
If we take (the real part of) $D$ and $d$ small enough to have UV convergence
even without the exponential factor, we can put $a=0$ and \reff{eq:P2} follows.

\vspace{0.5cm}
\noindent \textit{Property} 3.
\begin{equation} \label{eq:P3}
\sumprime{n}{p} \sumprime{m}{q}\ f(n,m) = 
 \sumprime{m}{q} \sumprime{n}{p}\ f(n,m) \quad .
\end{equation}

\vspace{0.5cm}
\textit{Proof}. \hspace{0.05cm}  
We give the proof only for the simpler case in which the function $f(n^2,m^2)$ does not
depend on the product $(m\cdot n)$.
Proceeding as in the previous case, one can introduce the auxiliary function
\begin{equation*}
f(a,n^2,m^2)=f(n^2,m^2)\ e^{-a(n^2+m^2)} \quad ,
\end{equation*}
for which the right- and left-hand side of~\reff{eq:P3} become
\begin{subequations} \label{eq:side2}
\begin{align}
\text{l.h.s. :} & \qquad \frac{1}{(2\pi i)^2} \int_{\Gamma_s} ds\ \zeta(s,p)\
 \int_0^\infty \!\! dx\ x^{s-1} \int_{\Gamma_u} du\ \zeta(u,q) \int_0^\infty \!\! dy\ 
 y^{u-1}\ f(a,x,y)  \tag{\ref{eq:side2}L} \\
\text{r.h.s. :} & \qquad \frac{1}{(2\pi i)^2} \int_{\Gamma_u} du\ \zeta(u,q)\
 \int_0^\infty \!\! dy\ y^{u-1} \int_{\Gamma_s} ds\ \zeta(s,p) \int_0^\infty \!\! dx\ 
 x^{s-1}\ f(a,x,y) \tag{\ref{eq:side2}R}
\end{align}
\end{subequations}
Again, both series have been defined without recourse to subtraction, by choosing
a large enough value for $\R s$ and $\R u$ along the contours
$\Gamma_s=\{\R s=c; c>p/2 \}$, $\Gamma_s=\{\R u=c; c>q/2 \}$.
Property 3 follows trivially for the auxiliary function by simply exchanging the
various integrals and we recover~\reff{eq:P3} by continuing both
sides (\ref{eq:side2}L), (\ref{eq:side2}R) analytically to $a=0$.



\begin{thebibliography}{99}
%
\bibitem{DR}
G.~'t~Hooft and M.~Veltman, {\it Nucl. Phys.} {\bf B44} (1972) 189. \\
 C.~G.~Bollini and J.~J.~Giambiagi, {\it Phys. Lett.} {\bf 40B} (1972) 566.
%
\bibitem{Wilson} K.~G.~Wilson, {\it Phys. Rev.} {\bf D7} (1973) 2911.
%
\bibitem{Collins} J.~C.~Collins, {\it Renormalization}, Cambridge University Press, 1985,
 Chapter 4.
%
\bibitem{statisticalM} In finite-size scaling theory, see: \\
E.~Br\'ezin and J.~Zinn-Justin, {\it Nucl.~Phys.}~{\bf B257}~[FS14]~(1985) 867.\\
J.~Rudnick, H.~Guo and D.~Jasnow, {\it J.~Stat.~Phys.}~{\bf 41}~(1985) 353.\\
J.~L. Cardy, editor, {\it Finite-Size Scaling}, North-Holland, Amsterdam, 1988.\\
J.~Zinn-Justin, {\it Quantum Field Theory and Critical Phenomena}, 3rd edition, 
 Oxford University Press, New York, 1996, 
 Chapter~36, in particular Appendix~A36.2 and references therein. \\
See also: S.~Caracciolo, A.~Gambassi, M.~Gubinelli and A.~Pelissetto, 
 {\it Eur.~Phys.~J.} {\bf B20} (2001) 255.
%
\bibitem{finiteT} In finite temperature field theory, see:  \\
 J.~Zinn-Justin, \texttt{hep-ph/0005272}.\\
 M.~Le~Bellac, {\it Thermal Field Theory}, Cambridge University Press, 1996.\\
 M.~Quiros, \texttt{hep-ph/9901312}. \\
 See also Ref.~\cite{Bedingham}. 
%
\bibitem{KK} In theories with compact dimensions, see: \\
 T. Appelquist and A. Chodos, {\it Phys. Rev. Lett.} {\bf 50} (1983) 141. \\
 T. Appelquist and A. Chodos, {\it Phys. Rev.} {\bf D28} (1983) 772. \\
 I.~Antoniadis and M.~Quiros,  {\it Nucl.~Phys.} {\bf B505} (1997) 109. \\
 I.~Antoniadis, S.~Dimopoulos, A.~Pomarol and M.~Quiros, 
 {\it Nucl.~Phys.} {\bf B544} (1999) 503.
%
\bibitem{effCas} For reviews on the Casimir effect see: \\
E.~M.~Santangelo, \texttt{hep-th/0104025}.\\
M.~Bordag, U.~Mohideen and V.~M.~Mostepanenko, {\it Phys.~Rep.}~{\bf 353}~(2001) 1.
%
\bibitem{zfunction} For $\zeta$-function regularization see: \\ 
 J.~S.~Dowker and R.~Critchley, {\it Phys.~Rev.} {\bf D13} (1976) 3224. \\
 S.~W.~Hawking, {\it Commun.~Math.~Phys.} {\bf 55} (1977) 133. \\
 E.~Elizalde, in Leipzig 1992, Proceedings,
 {\it Quantum Field Theory under the Influence of External Conditions}. \\
 E.~Elizalde, {\it Ten Physical Applications of Spectral Zeta Functions},
 Lecture Notes in Physics M~35 (1995) 1, Springer.
%
\bibitem{Ogreid} 
O.~M.~Ogreid and P.~Osland, \texttt{math-ph/0010026}.
%
\bibitem{GravitonLoops}
 R.~Contino, L.~Pilo, R.~Rattazzi and A.~Strumia, 
 {\it JHEP}~{\bf 06}~(2001)~005 [{\tt hep-ph/0103104}], Appendix~D.
%
\bibitem{Nibbelink} S.~G.~Nibbelink, \texttt{hep-th/0108185}.
%
\bibitem{Bedingham} D.~J.~Bedingham, \texttt{hep-ph/0011012}.
%
\bibitem{BHN} R.~Barbieri, L.~Hall and Y.~Nomura, {\it Phys. Rev.} {\bf D63} (2001) 105007.
%
\bibitem{NG}  M.~Ghilencea and H.~P.~Nilles, {\it Phys. Lett.} {\bf B507} (2001) 327.
%
\bibitem{Quiros} A.~Delgado, G.~von~Gersdorff, P.~John and M.~Quiros, 
  {\it Phys. Lett.}~{\bf B517}~(2001)~445.
%
\bibitem{CP}  R.~Contino and L.~Pilo, {\it Phys. Lett.} {\bf B523} (2001) 347.
%
\bibitem{Davies} B.~Davies, {\it Integral Transform and Their Applications}, 
 Applied Mathematical Sciences 25, Springer Verlag, Berlin, 1978.
%
\bibitem{GraRyz} I.~S.~Gradshteyn and I.~M.~Ryzhik, 
 {\it Table of Integrals, Series, and Products}, Alan Jeffrey ed., 5th edition, 
 Academic Press, New~York, 1994.
%
\bibitem{luscher} M.~L\"uscher and P.~Weisz, {\it Nucl.~Phys.} {\bf B445} (1995) 429.
%
\bibitem{CW} P.~Candelas and S.~Weinberg, {\it Nucl.~Phys.} {\bf B237} (1984) 397. 




\end{thebibliography}
\end{document}